\documentclass[11pt]{article}
\usepackage{epsf}

\def\1ad{\mbox{\normalsize $^1$}}
\def\2ad{\mbox{\normalsize $^2$}}
\def\3ad{\mbox{\normalsize $^3$}}
\def\4ad{\mbox{\normalsize $^4$}}
\def\5ad{\mbox{\normalsize $^5$}}
\def\6ad{\mbox{\normalsize $^6$}}
\def\7ad{\mbox{\normalsize $^7$}}
\def\8ad{\mbox{\normalsize $^8$}}

\def\npb#1#2#3{{\it Nucl. Phys.} {\bf B#1} (#2) #3 }
\def\plb#1#2#3{{\it Phys. Lett.} {\bf B#1} (#2) #3 }
\def\prd#1#2#3{{\it Phys. Rev. } {\bf D#1} (#2) #3 }

\def\bb#1{{\tt hep-th/#1}}

\def\jhep#1#2#3{{\it J. High Energy Phys.} {\bf #1} (#2) #3 }
\def\atmp#1#2#3{{\it Adv. Theor. Math. Phys.} {\bf #1} (#2) #3 }
\def\bb#1{{\tt hep-th/#1}}

\def\CN{{\cal N}}

\def\dj{\hbox{d\kern-0.347em \vrule width 0.3em height 1.252ex depth
-1.21ex \kern 0.051em}}

\def\Tr{{\rm Tr\,}}

\def\pt{\partial}

\newcommand{\be}{\begin{equation}}
\newcommand{\ee}{\end{equation}}
\newcommand{\ba}{\begin{eqnarray}}
\newcommand{\ea}{\end{eqnarray}}

\newcommand{\ie}{\mbox{{\em i.e.~}}}

\newcommand{\ben}{\begin{equation*}}
\newcommand{\een}{\end{equation*}}
\newcommand{\ban}{\begin{eqnarray*}}
\newcommand{\ean}{\end{eqnarray*}}
\newcommand{\brr}{\begin{array}}
\newcommand{\err}{\end{array}}
\newcommand{\bc}{\begin{center}}
\newcommand{\ec}{\end{center}}

\begin{document}

\newcommand{\sheptitle}
{On $1/N$ Corrections to the Entropy of Noncommutative Yang--Mills Theories}
\newcommand{\shepauthora}
{{\sc
J.L.F.~Barb\'on}}


\newcommand{\shepaddressa}
{\sl
Departamento de F\'{\i}sica
de Part\'{\i}culas. Universidad de Santiago \\
 E-15706 Santiago de Compostela, Spain \\
{\tt barbon@mail.cern.ch}}

\newcommand{\shepauthorb}
{\sc
E.~Rabinovici}

\newcommand{\shepaddressb}
{\sl
Racah Institute of Physics, The Hebrew University, Jerusalem 91904, Israel
 \\
{\tt eliezer@vms.huji.ac.il}}

\newcommand{\shepabstract}
{ We study thermodynamical aspects of string theory in the limit in which
it corresponds to Noncommutative Yang-Mills. We confirm, using the AdS/CFT
 correspondence,  that for general D$p$ branes
the entropy in the planar
approximation depends neither on the value of the background magnetic
field $B$ nor on its rank. We  find    $1/N^2$ corrections to the
planar entropy in the WKB approximation.
 For all appropriate values of $p$ these corrections are much softer
than the corresponding corrections for the $B=0$ case, and vanish altogether
in the high temperature limit.  
}

\begin{titlepage}
\begin{flushright}
US-FT/21-99\\
RI/15-99\\
{\tt hep-th/9910019}\\

\end{flushright}
\vspace{1in}
\begin{center}
{\large{\bf \sheptitle}}
\bigskip\bigskip \\ \shepauthora \\ \mbox{} \\ {\it \shepaddressa} \\
\bigskip\bigskip  \shepauthorb \\ \mbox{} \\ {\it \shepaddressb} \\
\vspace{0.5in}

{\bf Abstract} \bigskip \end{center} \setcounter{page}{0}
\shepabstract
\vspace{1in}
\begin{flushleft}
September 1999
\end{flushleft}


\end{titlepage}

\newpage

%
%

\section{\sf Introduction}
\setcounter{equation}{0}

In string theory, the perturbative two-dimensional  
world-sheet data contains information about the
target-space geometry background on which the string
propagates. Perturbatively one expects that an effective target-space
theory will have  on its world volume a definite geometry, the 
very same as that encoded
in the two-dimensional world-sheet theory.
 However,  one knows  from $T$-duality
that the metric on that world-volume may  actually be ambiguous \cite{trev}.  
 Moreover
there is no {\it a priori}  reason why  new ambiguities may not emerge
also non-perturbatively. Actually in some topological world-sheet
theories
the
topology of the target space in the sigma-model does not define
unambiguously the topology of
the world-volume in target space \cite{efr}. 
 In other cases such as type IIA string
theory  the  world-volume dimensionality of the effective target-space
theory is larger than ten and is actually eleven \cite{mth}. 

 Recently \cite{malda, gkpw}  it was
found that the world-volume of the effective target-space theory
may be smaller than that expected from the two-dimensional sigma-model,
 moreover the 
effective target-space theory may   be just a field theory not containing
gravity.
 For example  $\CN=4$, $D=4$, supersymmetric Yang--Mills (SYM)
 describes a string
propagating in the ten-dimensional target space  $AdS_{5} \times {\bf S}^5$.
 Not only
are the geometry, the dimensionality and topology of the target-space manifolds
largely modified but also the commutativity properties of the
target-space coordinates were found to be ambiguous: in a certain limit 
\cite{SW}, 
in which the background field $B$ is non-vanishing in some directions,
the metric components are of  $O(\epsilon)$ and the string Regge slope
 $
\alpha'$ is
of order $O(\sqrt{\epsilon})$, a 
string probe will not distinguish between one gauge system on a target-space
manifold whose variables are commutative  and
another gauge system on a noncommutative (NC) target space  whose
coordinates satisfy \cite{conn}
\be
[x^i , x^j] = i \,\theta^{ij}, 
\ee
with $\theta \sim B^{-1}$. 
This limit was originally introduced \cite{cds} as a limit of $M$-theory in
the context of the matrix model of ref. \cite{mmodel}.  The weak-coupling
picture in terms of standard open strings on D-branes was presented in
\cite{dhull} (see also \cite{li}) and further studied  
  from various points of view in \cite{bunch}. 

 For many years one has been curious about the manner in which
such non-commutativity would influence the behaviour of the
system. Given that in  string theory there are equivalences between
theories on commutative and non commutative spaces it may have not been
a total surprise that the entropies of both systems were found to be
indentical in a certain large $N$ limit. In fact it was shown that, at
the level of  
the large $N$ diagrammatic combinatorics,
 the modifications implied by the presence of the
torsion can be relegated to torsion-dependent phase factors mutiplying
torsionless $n$-point functions in momentum space \cite{comb, suskbig}. 
 In particular 
zero-point functions such as the free energy and the entropy are
unchanged. Although the  close similarity of thermodynamical functions 
 at low  temperatures (compared to  the
noncommutativity scale) was  expected, the fact that similarities
 persist, for large $N$, at high temperatures 
 was   more of a surprise. These identities  have
been demonstrated  also by applying the AdS/CFT correspondence and
considering the supergravity limit on the AdS side, a description valid
at strong 't Hooft coupling $g^2_{\rm YM} N \gg 1$. The appropriate
masterfield has been identified for the zero-temperature system \cite{HI, MR}
 and the 
additional masterfield, a black-hole configuration, has been identified
for the finite-temperature case \cite{MR}.
 While these supergravity configurations
do depend on the background torsion field, it has been shown by
calculating classically the black-hole entropy  that large $N$ thermodynamical
functions are  
 $B$-independent.

 The classical supergravity results
capture the leading  $N$ properties of  the gauge-theory side. 
There are no known limitations, based on diagrammatic analysis, 
 on the $B$-dependence of the entropy to
next to leading order in $1/N^2$. It may not be so easy to estimate these
corrections in the non-commutative gauge-theory language. However one may
also approach the calculation from the supergravity side of the 
correspondence. We
have used in ref. \cite{us} 
a WKB approximation to estimate the next to leading order
effects on the entropy for vanishing values of $B$. This method did not
take into account stringy quantum corrections but did account for
quantum  corrections to the classical supergravity picture. This is
actually usefull for the limit at hand in which stringy effects are
decoupled.

 As the main result of this paper, we
 find that the indroduction of the magnetic $B$-field, or
 in other variables  a non-infinite $ \theta$ background, leads to a
large  suppression of the next to leading order free energy and
entropy,
in fact the entropy correction vanishes at large temperatures,  
leaving only the leading order
entropy. Equivalently, the limit of infinite NC  length $\theta\rightarrow
\infty$, at {\it fixed} energy is completely saturated by the planar
limit.  For vanishing $\theta$, \ie for the gauge theory on a commuting
manifold the next to leading order results are of the same
qualitative properties as the leading result.

In section 2 we review the WKB approximation to the $1/N$ corrections to
the free energy and entropy. We apply the method to obtain highly
suppressed $1/N$ corrections in the case of the four-dimensional gauge
theory, \ie the case of a large stack of $N$ D3-branes. We also
recover the commuting case results for
 a vanishing value of $\theta$, that is in the limit of small and large
$B$ fields. 

 In section 3 these results are generalized to D$p$ branes (as well as
NS5-branes).
We study the entropy of the system as a function of both $p$ and the
rank $r$ of the magnetic field $B$. For all $p<7$ the  
large-temperature  planar entropy of
the black-hole configuration does not depend on the value of $B$. The
more complicated behaviour for $ 7>p>4 $ is  also  unchanged.
 The
 $1/N^2 $  
corrections do depend on the value of $B$. The main result is again that
we find  them to be  always
much softer than those corresponding to the $B=0$  case. We finally 
touch upon the most general configuration which supposedly describes the
supergravity side in the aditional presence of a constant electric field.

After this work has been completed we received the article \cite{oz}, which
contains some overlap with the material in section 3.

\section{\sf WKB Estimations of 1/N Corrections to the Entropy}

In the context of the AdS/CFT correspondence, the sum over diagrams
with toroidal topology in the 't Hooft classification should be
related, at strong coupling, to the one-loop diagrams of the corresponding
dual string theory. In our case,
we are interested in the one-loop
(toroidal world-sheet) diagram of type IIB string theory in the appropriate
string background  which leads to a  gauge theory.

Lacking an operative description in terms of an exact CFT, the calculation
of this diagram in closed form is beyond our capabilities. We can, however,
produce  estimates  by means of approximations to its low-energy limit:  
 the one-loop diagrams of IIB supergravity.

One observable which is rather independent  of  the regularization
 ambiguities of supergravity is the vacuum-subtracted
statistical free energy. At one-loop, the supergravity
fields can be regarded as free (interacting only with the background) and
the thermal free energy can be evaluated by means of an oscillator sum for
each field:
\be
\label{fen}
\beta\,F(\beta) = \sum_{\rm species} \Tr \,(-1)^{\rm F} \, {\rm log}\,
 \left(1-
(-1)^{\rm
F} \, e^{-\beta \,\omega} \right)
,\ee
where F is the space-time fermion number and the trace is over the
spectrum of physical fluctuations in the  background, with frequencies
$\omega$ given by the eigenvalues of the operator  $i\pt_t$, associated
to one particular temporal Killing vector of the background, which we assume
static. If we evaluate (\ref{fen}) by  some determinant on the compactified
 euclidean
continuation of the background, the
inverse temperature $\beta$ is the period of identification of the euclidean
time.                 

 The vacuum
energy has been subtracted in this definition of the free energy, so that
the relation to the one-loop path integral is
\be
I^{(1)}
 = \beta \,E_{\rm vac} + \beta\, F(\beta)
.\ee
Under the assumption of {\it local extensivity} we can estimate the
statistical sum by
\be
\beta \, F(\beta) \approx \int {\rm d}^d x \sqrt{|g|}
 \; \beta_{\rm loc} \, F_0 (\beta_{\rm loc})
,\ee
with a red-shifted local inverse temperature  $\beta_{\rm local} = \beta
\sqrt{g_{tt}}$, and $F_0$ the flat-space free energy. Namely, this is
an adiabatic approximation in which one assumes local thermal equilibrium
in cells which are still small compared to the global features of the
geometry, so that their contribution to the total free energy is given by
the red-shifted flat-space free energy of the cell, and one further assumes
extensivity with respect to the partition in cells.  These assumptions
can be justified with a standard application of the WKB approximation to
the solutions of the wave equation $\omega^2 + \pt_t^2 =0$ (see for example
refs. \cite{wkb}). Thus, the WKB approximation is good if the metric
is sufficiently smooth. For example, for $AdS_{d+1}$ with radius $R$, the
derivative of the inverse local temperature $|\pt_r \beta_{\rm loc}|$ is bounded
by $\beta/R$, with $\beta$ the temperature at the centre. According to the
AdS/CFT correspondence, this ratio is small precisely in the high-temperature
limit, and our WKB approximation is better the higher the temperature.   

For  massless fields in $d$ space-time dimensions, one has $\beta F_0 = -A
\,T^{d-1}$, with
 $A$ a constant proportional to the total number of particle degrees of
freedom.  
 Thus, the WKB approximation
involves integration of $(\sqrt{g_{tt}})^{1-d}$ over the volume. A more
geometrical characterization can be given  defining the {\it optical metric}
\cite{opticalm} by the conformal transformation
\be
ds^2_{\rm optical}  = {1\over g_{tt}} \,
 ds^2_{\rm euclidean}, 
\ee
from the Wick-rotated metric with compact time $t\equiv t + \beta$. 
Then, the contribution of a given region ${\bf X}$
of space-time
to the one-loop free energy is proportional to the optical volume of this
region, which we shall denote by ${\widetilde{\rm Vol}}({\bf X})$.

For space-times with a black hole, we should consider only the optical
volume of the region  not  excluded by the black-hole horizon. The reason being
that the euclidean metric of a black hole terminates at the horizon, which
represents a radial cut-off. 
 Strictly speaking, the red-shift estimate
breaks down very close to the horizon, because the local temperature diverges. 
 This local divergence can
be interpreted as contributing to the renormalization of the Newton constant,
as in ref. \cite{suskugl}. Thus, in dealing with black-hole spacetimes,
we shall consider only the optical volume of the asymptotic region,
sufficiently far from the horizon, \ie we consider the free energy of
the Hawking radiation in equilibrium with the black hole. This naive
subtraction of the horizon divergence is enough for the purpose of
order of magnitude  estimates \cite{us}. 

  For a curved space-time, a field is regarded as massless
if its mass is smaller than the local temperature $T_{\rm loc} =
T/\sqrt{g_{tt}}$. Otherwise, it is massive, and can be decoupled from the
statistical sum in that region of space-time. This means that, for a situation
with locally varying Kaluza--Klein thresholds one may partition the whole
manifold ${\bf X}$ in cells ${\bf X}_i$ of effective dimension $d_i$, defined
by the condition that the effective radii be sufficiently large compared to
the local temperature: 
\be
\beta_{\rm loc} \ll R_{\rm loc}
.\ee

  Finally,
 neglecting threshold effects, 
from the regions where $\beta_{\rm loc} \approx R_{\rm loc}$, 
  the WKB approximation to the   one-loop free energy can be written:
\be
I^{(1)}_{\rm WKB}
 = \beta \,E_{\rm vac} - \sum_i A_i \,T^{d_i} \,{\widetilde{\rm
Vol}} ({\bf X}_i)
,\ee
for a decomposition in ``cells" ${\bf X}_i$,
each of effective naive dimension $d_i$.

If we can isolate a regime where the dominant asymptotics is of the form
$\beta F \sim -A \, T^\gamma$, the corresponding one-loop entropy is
\be
S^{(1)}_{\rm WKB} \sim  (\gamma+1)\,A\,T^{\gamma}
.\ee
In this case, we can define
\be
d_{\rm eff} = \gamma+1
\ee
to be
 the effective dimension, as determined by the high-temperature asymptotics.
Notice that this dimension is in general different from the naive dimensions
$d_i$ of the cells in which we have partitioned the manifold. The reason
is that the optical volume of a given cell
 may depend non-trivially on the asymptotic
reference temperature. For example, for $AdS_{d+1}$,  and generalizations
involved in the AdS/CFT correspondence, the   effective dimension as
defined by the high-temperature asymptotics is $d$ instead of $d+1$,
\cite{us, usk}. This
is in fact a manifestation of holography at the level of $O(1/N^2)$ corrections.

\subsection{\sf The Basic Example}

The simplest example is given by the gravitational description
of the $\CN=4$ SYM theory at large $N$, obtained from a stack of D3-branes
with a non-zero
 $B$-field on a  single  spatial two-plane.
 At large 't Hooft coupling $\lambda = g_{\rm YM}^2 N$, the
master field of the theory with a NC
 parameter $
\theta$ is encoded in the metric derived in refs. \cite{HI} and
\cite{MR}:
\be\label{mblow}
{ds^2 \over \alpha' } =  {U^2 \over \sqrt{\lambda}} \left(
-dt^2 + dy^2 + {\hat f} (U)
\,d{\bf x}^2 \right) + \sqrt{\lambda} \left({dU^2 \over U^2} + d\Omega_5^2
\right)
\ee
with
\be
{\hat f} (U) = {1\over 1+ (U\Delta)^4}
\ee
and  $\Delta  = \lambda^{-1/4} \,\sqrt{\theta}$. Notice
that the perturbative NC energy scale, $\theta^{-1/2}$, differs
by powers of the 't Hooft coupling from the value of the  $U$-coordinate
threshold for the onset of NC effects in the metric (\ref{mblow}), which is given
by $U_\Delta = 1/\Delta$.  The associated length scale in the gauge theory,
according to the UV/IR correspondence  
  \cite{uvir}, is given by 
\be
a={\sqrt{\lambda}\over U_\Delta} = \lambda^{1/4} \, \sqrt{\theta}
,\ee
it differs from the weak-coupling NC  length 
scale, $\sqrt{\theta}$, by powers of
the 't Hooft coupling\footnote{  
We have absorbed various $O(1)$ constants in the definition
of $\lambda$ and $\alpha'$.}.  The small $U$ or infrared  region is the
standard $AdS_5 \times {\bf S}^5$  space
with radius $R= \sqrt{\alpha'} \,\lambda^{
1/4}$, in agreement with expectations, since NC effects should be irrelevant
in the deep infrared regime. Conversely, the $\theta \rightarrow 0$ limit
at fixed energy and coupling gives back the standard large $N$ master field
of the commutative theory. 

 In the non-extremal
case
the horizon sits at $U_0 = T \sqrt{\lambda} $. The local value of the inverse
temperature is $\beta_{\rm loc} = \beta \,U\,R /\sqrt{\lambda}
$ for $U\gg U_0$, while the local value of the ${\bf S}^5$ radius is
$R({\bf S}^5) = R$. So, for $U>U_0 =T\sqrt{\lambda}$ we
drop the five-sphere and the effective (euclidean)
 optical metric of interest is
\be
ds^2_{\rm optical}
 = dt^2 + dy^2 + {\hat f} (U) \,d{\bf x}^2  +
\lambda \,{dU^2 \over U^4}
,\ee
with optical volume
\be
{\widetilde{\rm Vol}} =\sqrt{\lambda} \int_{U_0}^{\infty}
 dt\,dy\,d{\bf x} \,dU \,{{\hat f} (U) \over U^2}
.\ee
So we finally obtain
\be
I^{(1)}_{\rm WKB} = \beta \,E_{\rm vac} - A \, (LT)^3 \,{\cal I} (aT)
\ee
in terms of the integral
\be
{\cal I} (aT) = \int_{1}^{\infty} {dx
\over x^2 (1+ (aT)^4 x^4)}
,\ee
which can be explicitly evaluated:
$$
{\cal I} = 1 -{\pi aT \over \sqrt{8}} + {aT \over 4\sqrt{2}} \left[ 2
\;{\rm arctan} \;(1+\sqrt{2} aT) -2\;{\rm arctan}\;(1-\sqrt{2}aT) + {\rm
log} \; \left({1-\sqrt{2} aT + (aT)^2 \over 1+ \sqrt{2} aT +
(aT)^2}\right)\right]
$$
The important feature of this function is that it represents a small
correction at low temperature $aT \ll 1$:
\be
{\cal I} = 1-{\pi aT \over \sqrt{8}} + {(aT)^4 \over 3} - {(aT)^8 \over
7} + \dots
,\ee
but a large suppression for very high temperatures, compared to the NC scale
 $aT \gg 1$:
\be
{\cal I} \longrightarrow {1\over 5 (aT)^4} - {1\over 9 (aT)^8} + \dots
\ee

This means that the one-loop free energy scales like a vacuum contribution
in the large temperature limit $aT \gg 1$. In other words, the
 $1/N^2$ corrections to entropy {\it vanish} in 
such a limit: 
the extra contribution from the extreme ultraviolet  
 regime is as if it represented a  
zero-dimensional volume.
As mentioned in the introduction, the planar $O(N^2)$
 entropy is expected to be the same as the non-commutative
one, on the grounds of  weak-coupling arguments  \cite{suskbig}. This fact was verified
at strong-coupling  in ref.  
\cite{MR}, \ie the horizon area in Einstein frame does not change beyond
the NC scale $a$. Our result indicates that, 
 at least to leading order,   the source
of all the
high-temperature entropy is in the planar evaluation of  degrees of freedom.  

Conversely, we can say that, in the limit of large NC parameter $\theta\rightarrow\infty$,
at fixed energy, we are led to a purely planar theory, in the sense that
the 
large $N$ description is effectively classical (trivial $1/N$ corrections).

\section{\sf Generalizations}

Most of the previous discussion  admits generalization
 to general D$p$-branes with $1<p<7$. The supergravity string-frame
solution for a stack of $N$ D$p$-branes with an aligned $B$-field
(before the decoupling limits)    is given  in ref. \cite{MR}: 
\be
\label{bbrane}
ds^2 = {1\over \sqrt{H(\rho)}}
 \left(-dt^2 + d{\bf y}^2 + f(\rho)\, d{\bf x}^2 \right)  + \sqrt{H(\rho)} 
\left(
d\rho^2 + \rho^2 d\Omega^2_{8-p} \right)
.\ee
There is a $B$-field of  rank $2r$
 in the  {\it spatial} directions ${\bf x}$ of intensity\footnote{We write here
the value of the skew-eigenvalues, which we assume all equal in magnitude, for
 simplicity of notation.} 
\be
\alpha' {\bar B} = {\rm tan}\,\vartheta \; {f(\rho) \over H(\rho)}
\ee
and dilaton
\be
e^{2\phi} = e^{2\phi_{\infty}}\, H(\rho)^{3-p \over 2} f(\rho)^r
.\ee
The functions $H(\rho)$ and $f(\rho)$ are given by
\be
H(\rho)= 1+ (R/\rho)^{7-p}
,\qquad
f(\rho)^{-1} = {\rm sin}^2 \,\vartheta \; H(\rho)^{-1} + {\rm cos}^2 \,
\vartheta
.\ee
The basic low-energy  scaling of ref. \cite{malda} $\rho= \alpha' U$ leads to
\be
H(U) \longrightarrow {\lambda \over  (\alpha')^2 U^{7-p}}
,\ee
where $R = \sqrt{\alpha'} \;(G_s N)^{1\over 7-p}$ in terms of the
NC string coupling $G_s$ and the corresponding  't Hooft
coupling
$
\lambda = g^2_{\rm YM} N = (\alpha')^{p-3 \over 2} \; G_s N
,$
where we have again absorbed  various constants of $O(1)$ into the definitions
of the parameters.

The low-energy scaling introduced by  Seiberg and Witten in ref. \cite{SW}
  shrinks
  the closed string metric in
 the direction of the ${\bf x}$ coordinates, with a constant $B$-field.
We may achieve this in the previous solution by a rescaling of
 the coordinates ${\bf x} \rightarrow \alpha' {\bf x} /{\theta}$, in
the $\alpha' \rightarrow 0$ limit 
 with constant ${\theta}= \alpha' \;{\rm tan}\,
\vartheta$.
At the same time, the original string coupling is scaled $e^{\phi_\infty}
 \rightarrow G_s \,
(\alpha' / {\theta})^r$, and the $B$-field transforms like a normal tensor
under the rescaling:
$
{\bar B} \rightarrow  B \cdot (\alpha' /{\theta})^2
.$
 In terms
 of the convenient NC length scale $\Delta$ defined by
\be
\Delta^{7-p} = {{\theta}^2 \over \lambda}
,\ee
we get the following
 string metric after this double Maldacena--Seiberg--Witten scaling:
\be\label{mfield}
{ds^2 \over \alpha'} = {U^{7-p \over 2} \over \sqrt{\lambda}} \left( -dt^2 + d{\bf
y}^2 + {\hat f} (U) \, d{\bf x}^2 \right) + {\sqrt{\lambda} \over  U^{7-p \over 2}}
\; \left( dU^2 + U^2 d\Omega^2_{8-p} \right)
,\ee
with
\be\label{efe}
{\hat f} (U) = {1\over 1+ (U \Delta   )^{7-p} }
.\ee
This result agrees with the recent determination of this function in
ref. \cite{lit}. The $U$-dependent $B$-field profile is
\be
 B =  B_{\infty} \; (U  \Delta )^{7-p} \;{\hat f} (U)
,\ee
with $ B_{\infty} = 1/{\theta}$. 
This asymptotic value of the $B$-field agrees with the zero slope limit
of ref. \cite{SW} for the NC parameter matrix:
\be
\theta^{ij} = 2\pi \alpha' \left( {1\over g + 2\pi \alpha'  B_\infty} \right)^{ij}_A
\ee
with $g_{ij}$ the closed-string metric.    
A potential confusion stems from the fact that the NC parameter in this formula
vanishes {\it both} for large and small values of the $B$-field. On the other hand,
if $B_{\infty} = 1/ \theta$, the limit of vanishing $B$-field seems to 
make NC effects blow up. This is resolved by noticing that the NC parameter
vanishes with the $B$-field {\it only} if $\alpha'$  and the closed-string metric are
 kept fixed, namely the two limits  that  turn-off $\theta$ do not commute. In the 
supergravity solution, keeping the open-string scale fixed (Born--Infeld corrections)
amounts to keeping the ``neck" of the throat at $U_s \sim (\lambda/(\alpha')^2)^{1\over
7-p}$ in place in the full solution (\ref{bbrane}). If we now take the vanishing $B$-field
limit with constant $g_{ij}$ and constant $\alpha'$, we find $f(\rho) \rightarrow 0$
and NC features vanish as it should be.

Coming back to the scaled solution, the dilaton is
\be
e^{2\phi} = G_s^2 \; {\hat f}(U)^r \; H(U)^{3-p \over 2}
 = e^{2\phi_{\rm C}} {\hat
f}(U)^r
,\ee
where  $\phi_{\rm C}$ denotes
 the dilaton of the $\Delta=0$ theory.

 With these data, one could
study the interplay of 
phase transitions in these models, depending on
the local duality transformations appropriate   for each description. 
 Compared
to the analysis of the  commutative case in ref. \cite{maldais},
 the NC character introduces a new scale in
the problem at $U_{\Delta} = 1/\Delta$, associated to the onset of NC effects.
The corresponding {\it length scale} in the gauge theory, according to the
generalized UV/IR correspondence of \cite{uvir} is
\be\label{coeq}
a =\sqrt{\lambda \over U_\Delta^{5-p}} = \sqrt{\lambda \,\Delta^{5-p}}
.\ee

Following \cite{maldais}, the  applicability of the
supergravity description is  controlled by the size of $\alpha'$ corrections
in the string-metric background. In terms of the ``correspondence point"
$U_c =\lambda^{1\over 3-p}$  of ref.  \cite{horpol},
 one finds that the geometric description
is good for $U\ll U_c$ when $p<3$. Therefore, we need $U_\Delta < U_c$ in
order to trust the supergravity solution in the region where NC effects
are sizeable.  In terms of the 't Hooft coupling versus the gauge-theory
 NC length scale,
 this condition is
$\lambda^{3-p} \,a < 1$, \ie we require a sufficently {\it weak}
 coupling. Otherwise, the NC features of the
ultraviolet regime must be studied entirely by means  of perturbative techniques.
 
For $p=3$, the condition for the supergravity picture to capture NC features
in the ultraviolet is the ordinary one,  independent of the  scale: $\lambda >1$.

On the other hand, for $p=4$ the supergravity patch is $U \gg U_c = 1/\lambda$,
so that the NC features are visible in the supergravity description for
sufficiently {\it strong} coupling: $\lambda > a$.

 Finally, 
 the cases $p=5,6$ are somewhat different since they do not
follow a standard IR/UV correspondence (equation (\ref{coeq}) does not have a
clear physical interpretation in these cases).
Still, we can associate NC effects to the {\it energy} scale $U_\Delta$, as measured
for example by the mass of a
 stretched fundamental string probe. The condition
for the metric (\ref{mfield}) to accurately describe the NC effects is thus
$\lambda \,(U_\Delta)^{p-3} >1$. 
  
The non-perturbative thresholds associated with large values of the
string dilaton are generally relaxed by turning on the NC moduli. Since
${\hat f} \rightarrow U^{p-7}$ vanishes in the large $U$ regime, this means
that the present metrics have small local string coupling in the $U\rightarrow
\infty$ region for {\it all} values of $p$,
 provided $r\geq 1$ (in fact, one needs the slightly stronger
condition $r\geq 2$ for $p=6$).
 Following the general rule, the infrared
 thresholds
associated with small $U<U_\Delta$ singularities are qualitatively
 the same in the NC case.

 In general, there could be intermediate regimes
with large local string coupling, but such transients can be ignored when
 working
in the 't Hooft limit with fixed values of the typical energies in the system,
as well as  $\lambda$ and $\Delta$, of $O(N^0)$.

\subsection{\sf Planar Thermodynamics}

The (somewhat surprising)   robustness of the
planar thermodynamics of the $p=3$ case, discussed in \cite{MR},
 persists for general values of $p$. The non-extremal
metric is obtained by replacing
\be\label{repl}
-dt^2 \rightarrow  + h\, dt^2,  \qquad dU^2 \rightarrow dU^2 / h
,\ee
with the euclidean time identified with period the inverse temperature $t
 \equiv t+ \beta$ and
\be\label{hache}
h= 1- (U_0 / U)^{7-p}
\ee
as usual. Since no $B$-field lies in the time direction, these replacements do not
affect the parts of the metric which depend on $\Delta$ (the ${\bf x}$ space).
Therefore, the NC Hawking temperature is the same as in the commutative
 case.
\be
T_{\rm NC} = T_{\rm C} ={7-p \over 4\pi} \sqrt{U_0^{5-p} \over \lambda}
.\ee

Moreover, the planar entropy is also independent of the NC deformation
parameter. Since it must be computed in  the Einstein frame,
we have to multiply the string metric by
$
e^{-\phi_{\rm NC}/2} = e^{-\phi_{\rm C} /2} {\hat f}(U)^{-r/4}
$.
The horizon being eight-dimensional, this
 yields a factor of $({\hat f}^{-r/8})^8
$, which exactly cancels the extra factor of $\left(\sqrt{\hat f}
\right)^{2r}$ coming from
the $2r$ directions with a non-vanishing  $B$-field. So, the NC horizon area is
\be
A_{\rm NC} = A_{\rm C} \; ({\hat f}^{1/2})^{2r} \; ({\hat f}^{-r/8})^8 = 
A_{\rm C} 
.\ee
Both the planar entropy and the temperature are exactly the same as in
the commutative case, which means that all {\it planar} 
 thermodynamical functions are the same.

\subsection{\sf   WKB Corrections to the Entropy}

In order to estimate the $1/N$ corrections, we consider the corresponding
 optical metric
\be
ds^2_{\rm optical}
 = dt^2 + d{\bf y}^2 + {\hat f} (U) \,d{\bf x}^2 + {\lambda \over
U^{7-p}} \; \left( dU^2 + U^2 d\Omega^2_{8-p} \right)
,\ee
and compute the optical volume of the region $U_0 < U <\infty$. 
The conditions for decoupling the angular sphere ${\bf S}^{8-p}$ are the
same as in the commutative case, again because the NC character only 
affects the
${\bf x}$--space. We discuss the qualitatively different cases in turn.

\vspace{0.5cm}

{\underline {\sl D$p$-branes with $ p<5$}}  

\vspace{0.5cm} 

For $p<5$, the temperature is small: 
$\beta_{\rm loc} > R({\bf S}^{8-p})_{\rm loc}$ in the region of
interest, so that  we can drop the angular sphere in  
estimating   the free energy of thermal radiation outside the black-brane.  
\be
I_{\rm WKB}^{(1)}  \sim -T^{p+2} \,
{\widetilde {\rm Vol}}_{p+2} = - T^{p+1} \,L^p
\int_{U_0}^{\infty} dU {\hat f}(U)^r \sqrt{\lambda U^{p-7}}
 \sim  - (L\,T)^p  \int_1^\infty
{dx \sqrt{x^{p-7}} \over (1+ (x U_0 \Delta  )^{7-p})^r}
.\ee
This is the standard result of the commutative theory $I_{\rm WKB}^{(1)} \sim -(L\,T)^p$
 for $U_0 \Delta
 \ll 1$. On the other hand,   in the opposite limit
 $T a \equiv T \sqrt{\lambda \Delta^{5-p}} \gg 1$,  we get a strong
suppression
\be
I_{\rm WKB}^{(1)}   \longrightarrow
 - {(L\,T)^p \over (a\, T)^{2r(7-p) \over 5-p}}
.\ee
Thus, we find
 the soft behaviour at high
 temperatures of the one-loop free energy, much like the D3-brane
 case. Notice that for all $p<5$, the asymptotic effective
exponent of $T$ is negative provided $r\geq 1$. Therefore,
 the effective dimensionality, as determined by
 the one-loop corrections, drops to zero or  is even ``negative'' 
at $T a \gg 1$.

\vspace{0.5cm}

{\underline {\sl D5-branes}}

\vspace{0.5cm}

For $p=5$ one gets, independently of the issue of angular decoupling:
\be
I_{\rm WKB}^{(1)}
 \sim -(L\, T)^5 \int_1^\infty {dx \over x} {1 \over (1+ (x U_0 \Delta  )^2
)^r}
.\ee
As long as $r > 0$, the integral converges! This is an 
improvement with respect to the commutative
 case, with $\Delta =0$, in which one gets a logarithmic divergence
of dubious interpretation. In the $U_0 \Delta  \gg 1$ regime one finds
\be
I_{\rm WKB}^{(1)}  
 \longrightarrow -{(L\,T)^5 \over (U_0 \Delta  )^{2r} }
.\ee
However, now the energy-density
 parameter $U_0$ is unrelated to the temperature,
which is constant and equal to $\lambda^{-1/2}$, \ie  there is a suppression
of the non-planar corrections, although the effective dimension  remains
$d_{\rm eff} = 6$.

\vspace{0.5cm}

{\underline {\sl D6-branes}}

\vspace{0.5cm}

On the other hand, for $p=6$, the local temperature outside the horizon is higher
than the mass of angular modes and  we must consider the optical volume of the
 angular sphere ${\bf S}^{8-p}$ as well. The
resulting one-loop free energy is
\be
I_{\rm WKB}^{(1)}  \sim -(L\,T)^6 \int_1^\infty {dx \sqrt{x} \over ( 1 + x U_0
\Delta  )^r}
.\ee
Now, we need a $B$-field turned on in at least two  planes ($r>1$), in order to achieve
convergence at large $U$ (this is reminiscent of the analogous condition to have a
vanishing string coupling at infinity). In any case, the interpretation is not clear, because
the standard UV/IR relation breaks down at the level of the formula for the
Hawking temperature, since large energies (large $U_0$),
 correspond to low temperatures. In fact, the  scaling at
 large temperature is that of
a higher-than-seven-dimensional theory:
\be
I_{\rm WKB}^{(1)}
 \sim -{(L\,T)^6 \over (U_0 \Delta )^r} \sim - L^6 \, a^{2r} \, T^{6+2r}
.\ee

Therefore, the cases $p=5,6$ continue to have non-standard features, although we
do see a general tendency of the $B$-fields to make the large $U$ behaviour
less singular in all cases.

\vspace{0.5cm}

{\underline {\sl Other Models} }  

\vspace{0.5cm}

These WKB estimates can be extended to other interesting models. For example,
we may consider NS5-branes of type IIB and IIA  related to D5-branes by a sequence of
$S$- and $T$-dualities. The behaviour of $1/N^2$ corrections for all these models
is essentially equivalent to that of type IIB D5-branes, \ie the commutative versions
have semi-infinite  cylinders that produce logarithmically divergent Hawking-radiation
entropies \cite{malstro, us, usk}. On the other hand, turning on $B$-fields regulates this divergence and
implies an effective quenching of $1/N$ corrections at large temperature. This is particularly clear
for the case of type IIB NS5-branes, whose metric is $S$-dual to that of D5-branes.
Since this duality amounts to a conformal transformation of the metric, to which
 the optical
metric is insensitive, we get the same physics of
$1/N^2$ corrections: for large energy densities $\Delta U_0 \gg 1$,  
\be
\left[{I_{\rm NC} \over I_{\rm free\;gas}}\right]^{(1)}_{\rm IIB \;NS5}
 \longrightarrow (U_0 \Delta)^{-2r}
.\ee

Type IIA NS5-branes can be obtained from type IIB NS5-branes by a further $T$-duality
along a commutative direction.  We have a global factor of ${\hat f}^{-r/2}$
 from
the $S$-duality transformation from D5-branes to type IIB NS5-branes. $T$-duality
inverts this factor on one of the commuting coordinates. In addition, there are
the usual factors of ${\hat f}$ for each NC coordinate. Thus, the optical volume
integrand gets an additional factor of ${\hat f}^{r/2}$ in all, leading to  
\be
\left[{I_{\rm NC} \over I_{\rm free\;gas}}\right]^{(1)}_{\rm IIA \; NS5}
 \longrightarrow (U_0 \Delta)^{-3r}
,\ee
again in the large density limit.
 When the type IIA D4-brane solution is lifted to
eleven dimensions, one obtains a NC 
 M5-brane model with $AdS_7 \times {\bf S}^4$
geometry in the infrared region. The previous scaling gives now 
\be
\left[{I_{\rm NC} \over I_{\rm C}}\right]^{(1)}_{\rm M5}
 \longrightarrow (U_0 \Delta)^{-9r/2}
\sim (a\,T)^{-9r}.
\ee

 We remark that 
one interesting  case was  not discussed here in detail. It is the case
in the presence of a 
 nonvanishing ``electric" NS-fields: $B_{0i} \neq 0$. Naively,
one expects similar results to the purely  ``magnetic" case,
at least as far as the arguments of ref. \cite{suskbig} concern. However,
it was pointed out in ref. \cite{MR}  that, at least in the particular case
of $p=3, r=2$, the supergravity picture of the
thermodynamics is fundamentally different at temperatures of the order
of the timelike noncommutative scale.

Assuming that, for $B_{0i}\neq 0$,
 the dominant finite-temperature master field (the black hole) is also
given by the substitutions (\ref{repl}) and (\ref{hache}) on the extremal
solution, one finds that the behaviour described in \cite{MR} for
$p=3$ actually generalizes to all D$p$-branes (this assumption might
actually hide important subtleties, related to the proper treatment of the
Wick rotation). 
 The perturbative  scale of ``electric"
non-locality is $\sqrt{\theta_e} = 1/\sqrt{B_{0i}}$. At large 't Hooft
coupling it develops into the length scales
\be
\Delta_e = \left({\theta_e^{\,2} \over \lambda}\right)^{1\over 7-p}, \qquad
a_e = \sqrt{\lambda \, \Delta_e^{5-p}}, 
\ee
that characterize the supergravity solution. The temperature is related
to the horizon radius $U_0$ by $T= T_{\rm C} (U_0) \, \sqrt{{\hat f}_e
 (U_0)}$,
where $T_{\rm C}$ is the temperature of the commutative theory and  
${\hat f}_e (U)$ is given by eq. (\ref{efe}) upon
 replacing $\Delta$ by $\Delta_e$. This temperature/mass relation
leads to negative specific heat for $U_0 \gg 1/\Delta_e$ and a maximum
temperature of order $T_{\rm max} \sim 1/a_e$, a behaviour reminiscent 
of standard black-branes in asymptotically flat space,  
 before the near-horizon scaling is taken as in eq.
 (\ref{bbrane}).
In fact,  the Einstein-frame metrics are  exactly equal to the string-frame
metrics of such asymptotically flat D-brane metrics  (up to
some rescalings of the coordinates,)  precisely if $2r=p+1$,
\ie when all the world-volume of the brane is noncommutative. This property
was noticed in ref. \cite{MR} for the $p=3, r=2$ case. However, we
see that the important qualitative features (a maximum temperature 
 and a negative specific heat branch at large energy
densities) generalize for arbitrary values of $p$ and $r$,
provided the {\it time} direction is noncommutative.

Having asymptotically flat regions at large $U$ will surely complicate
the workings of holography in these models. In particular, our WKB estimate
for the one-loop entropy gives a {\it
 ten-dimensional} contribution, $S^{(1)} \sim
T^9$,  from
supergravity modes at large $U\gg 1/\Delta_e$ in these models. 
Perhaps the theory imposes an effective cut-off of order $U_{\rm max} \sim 1/
\Delta_e$ already at the planar level, as suggested by the existence of
a maximum temperature at this scale. In this respect, it is interesting
to notice that the branch with negative specific heat at large energy
densities is dynamically suppressed in the canonical ensemble. The planar
 entropy
in these models is given by  $S = {\hat f}_e (U_0)^{-1/2} \, S_{\rm C}$, with
$S_{\rm C}$ the entropy function of the commutative theory. Using the $T(U_0)$ 
function one finds that the entropy scales 
as $S \propto N^2 \, T^{p-8}$ in the negative specific
heat branch. From here one can get the free-energy excess over the vacuum: 
\be
(I -\beta\;E_{\rm vac})_{\rm planar} \longrightarrow + {C_p \over 7-p}\,
 {N^2 L^p \over \lambda^{11-p \over 2}}
 \; \Delta^{-(p-7)^2 /2} \, (\beta_{\rm NC})^{8-p} >0
,\ee
with $C_p >0$. It is {\it positive} in the region with negative specific heat.
 Therefore, there is
an $O(e^{-N^2})$ suppression of this unstable branch in the canonical ensemble.

Another aspect of these solutions that we did not analyze in detail is the
presence of light thermal winding modes at large radial coordinates. These
cannot be eliminated through $T$-duality, and are sure to affect the
physics at large values of $U$.

\newpage 

\section*{\sc Acknowledgements}

J.L.F.B.    would like to thank A. Gonz\'alez-Arroyo and C. G\'omez for
 useful discussions and hospitality at
 the ``Instituto de F\'{\i}sica Te\'orica,
C--XVI, Universidad Aut\'onoma de Madrid,"
	   where part of this
work was carried out, as well as to the Spinoza Institute, University of
Utrecht, where this work was initiated.  

E.R. would like to thank A. Hashimoto,  N. Itzhaki and N. Seiberg
 for discussions, 
  the ``ITP Program on Gauge Theories and String Theory"
 at Santa Barbara, 
and the Randall Laboratory of Physics at the  University of Michigan,  
 where part of this work was done, for providing  stimulating enviroments. 
The research of 
E.R. is  partially supported by the BSF-American Israeli Bi-National
Science Foundation and the IRF Centres of Excelency Program.

%

\end{document}